# Chatbot integration in few patterns

**Marcos Baez**
Université Claude Bernard Lyon 1, Lyon, France

**Florian Daniel**
Politecnico di Milano, Milan, Italy

**Fabio Casati**
Tomsk Polytechnic University, Russia

**Boualem Benatallah**
University of New South Wales, Sydney, Australia

Chatbots are software agents that are able to interact with humans in natural language. Their intuitive interaction paradigm is expected to significantly reshape the software landscape of tomorrow, while already today chatbots are invading a multitude of scenarios and contexts. This article takes a developer's perspective, identifies a set of architectural patterns that capture different chatbot integration scenarios, and reviews state-of-the-art development aids.

*This paper was written with the precious contribution of co-author Florian Daniel, who passed away a few days after completing this manuscript for submission. We remember him here.*[1]

According to Gartner (https://gtnr.it/2MHVDG3, accessed April 1, 2020), by 2020 "twenty-five percent of customer service and support operations will integrate virtual customer assistant (VCA) or chatbot technology across engagement channels," while according to data from Statista.com the number of "digital assistants [users] worldwide is projected to reach 1.8 billion by 2021" (https://bit.ly/2HZlOZM, accessed April 1, 2020). As a matter of fact, chatbots have found their way into our everyday life without creating much discomfort: through platforms such as WhatsApp, Facebook Messenger and WeChat that can each enable conversational access to services to more than a billion of monthly users (https://bit.ly/2o3eKlb, accessed April 1, 2020); while digital personal assistants like Amazon Alexa, Apple Siri and Google Assistant are opening up new markets for voice users (Amazon alone has already sold more than 100 million Echo devices, https://bit.ly/2KnTD8w, accessed April 1, 2020).

The efficient development of chatbots – both written and spoken – is thus getting crucial to cope with the expected growth. While building robust intelligent chatbots is still a challenging endeavor, a myriad of well-thought and easy-to-use development frameworks have emerged to support the full life cycle of chatbots from natural language processing to invoking application programming interfaces[2]. As development support matures, the challenge is now shifting to understand how to integrate chatbots seamlessly into existing IT systems, knowledge bases, and business practices. That is, the questions are which vocabulary and intents should the bot master, which types of actions should it support, and how to enact them on a pre-existing software system.

So far, chatbot development has been studied considering different aspects of design, such as interaction model, application domain, goal-orientation, and dialog management[3]. Existing surveys have analyzed popular chatbot systems and chatbot frameworks (e.g., Harms et al.[2]). All these classifications follow a white-box approach and focus on the ingredients that define the internals of a chatbot, which translate into conversational capabilities and, eventually, user experience.

We propose an original perspective on chatbot development - an architectural gray-box perspective - and highlight fundamental differences in concepts, technology and purpose across existing chatbots. Here, we use the term chatbot in its broader sense, to refer to conversational agents of any kind enabling conversational access to software-enabled services. For example, a chatbot providing conversational access to a database may enable users to search and navigate data schemas translating user inputs into SQL queries, while a chatbot providing in-app assistance to users of an e-commerce website may feature guidance on product search or checkout options by highlighting HTML elements in the website. Where a chatbot is integrated into an existing system, e.g., into database or graphical user interface, determines how conversations must be structured and how intents, chatbot logic and actions must be configured.

Reasoning on the traditional three-layered architecture of applications, our own experience, and a systematic analysis of 347 papers reporting on chatbot systems in the last five years, we identified eight patterns that express distinct scenarios of how to integrate a chatbot into existing software systems. In this paper, we complement these eight resulting patterns with pointers to respective development aids that are particularly relevant to researchers, software architects and developers respectively looking for novel research domains and reusable integration knowledge.

# CHATBOT DESIGN DIMENSIONS

There are many approaches to the development of chatbots[3], and the choice relies on the type of service and experience the developer plans to deliver to its users. From this perspective, there are generally two categories of chatbots, i) *task-oriented*, which are designed to serve specific tasks in a specific domain, e.g., a weather chatbot, and ii) *chit-chat bots*, which tend to serve no specific purpose but aim at holding open-domain conversations with users. Modern task-oriented chatbots are built on a frame-based architecture, which relies on a domain ontology (composed of frame, slots and values) that specify the type of user intentions the system can recognize and respond to[2].

The tasks to be served and the complexity of the style of conversation shape the definition of intents, actions and the dialog control. *Intents* are the conceptual requests by the user, i.e., the tasks to be performed. They are provided in natural language through so-called *utterances*, where one or more utterances may express the same intent. Identifying user intents (e.g., obtain a weather forecast) from utterances (e.g., "What's the weather like today?") requires a *natural language processing unit* (NLP). In order for the NLP unit to know how to map utterances to intents, it is *trained* with a dataset of examples of utterance-intent mappings. Intents may have parameters, so-called *slots* (e.g., the date of a weather forecast), and the language understanding part of the NLP must be able to infer their *values* from utterances (e.g., date: today). Once an intent is identified, the *dialog management* component enacts an appropriate *action*, i.e., a specific operation serving the intent (e.g., perform a call to the weather API). To disambiguate similar intents or infer values of slots, additional information, the dialog *context*, may be used (e.g., if the chatbot already knows the location of the user, it does not need to ask for it in order to provide a localized weather forecast). The dialog control is designed either explicitly by defining conversation flows or derived from previous conversations, or using a combination of both techniques (refer to Harms et al.[2] and Hussain et al.[3] for more details on chatbots design and architecture). We illustrate these concepts in Figure 1a.

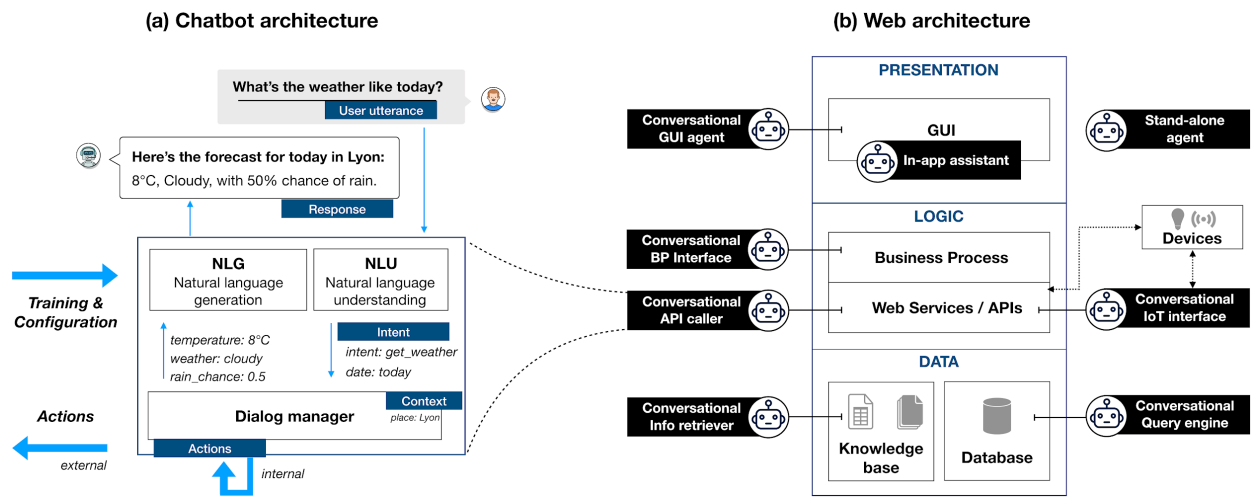

Figure 1. (a) Conceptual architecture of a chatbot. (b) Chatbot integration patterns in reference to the typical three-tier architecture of Web-based systems. The tier where the chatbot is located steers the training and configuration of the bot, the conversation flow between the bot and the user, and the resulting actions.

# CHATBOT INTEGRATION

The problem of integrating software has generally been formulated in function of what exactly is to be integrated. We commonly distinguish between data integration, application integration and user interface (UI) integration, in line with the typical three-tier architecture of distributed software.[4] *Data integration* brings together data schemas and data from different sources using techniques like schema mapping and entity resolution. *Application integration* connects software systems through their APIs or backend services, e.g., using software adapters and service orchestration, or through their UIs, like in robotic process automation[5]. *UI integration* connects applications by rendering together independent widgets or pieces of UIs that may have their own application logic and data, using for example HTML templates and in-browser event propagation for synchronization.

We define *chatbot integration* as the problem of integrating conversational capabilities into existing software systems. Doing so may require developing a conversational agent that starts from either the data, application logic or graphical UI of the system to support natural language conversations leveraging Artificial Intelligence (AI) or more traditional software engineering approaches.

Figure 1b illustrates the eight patterns for chatbot integration we identify in relation to the traditional three-layered architecture of distributed systems. Taking this architecture as a reference point, we derived the integration patterns considering the following criteria i) to what layer of the architecture is the chatbot integrated (e.g., Application logic), and what specific component within the chatbot acts upon (e.g., API, Business process), and ii) how uniquely the layer and related components inform the chatbot design dimensions and capabilities.

We followed a mixed approach in deriving and refining the integration patterns. We first identified a set of relevant papers, which were discussed in two iterations to reach an initial set of seven patterns. In order to validate and refine this original set, we then performed a systematic search on Elsevier's Scopus database for papers focusing on *chatbots* (keywords: chatbot, talkbot, conversational agent, voice user interface, smart speaker, smart assistant, Amazon Alexa, Google Assistant), describing *implementations* (keywords: implementation, prototype, system, architecture), and published in English language since 2015 until March 5, 2020. The resulting 938 papers were screened by two researchers. In an initial phase a sample of 100 papers were screened and discussed by both researchers (coding agreement 87%), and the rest was divided and annotated independently. In the process, the researchers annotated each paper based on i) relevance, ii) the associated pattern (according to the pre-defined criteria), while taking note of iii) potential deviations from the initial patterns. These deviations were discussed jointly by the researchers, and as a result some of the existing patterns were refined (definition and scope) and one new pattern emerged, for a total of eight integration patterns.

In the next section, we elaborate each of the identified patterns in more detail, concentrating on the core differences in setting up the respective chatbots. We recall from the design dimensions that configuring a chatbot requires developers to provide (i) the *intents* the chatbot should understand, (ii) *training* data to instruct the NLU, (iii) *action* implementations to serve intents, and (iv) dialog control implementation to manage the conversation. Thus, we focus on these four aspects to describe how the characteristics of the integration patterns shape the chatbot design dimensions and inform the development support provided by frameworks and platforms.

# INTEGRATION PATTERNS

To illustrate the extent of the research on each of the integration patterns, we start by briefly describing the results of the systematic screening process. We identified 347 relevant chatbot systems, with 290 referring to specific chatbot implementations and 64 to frameworks, platforms or methods supporting chatbot development. The resulting integration patterns illustrated in Figure 1b are *stand-alone* (240), *information retriever* (65), *IoT interface* (21), *query engine* (9), *GUI agent* (4), *in-app assistant* (4), business process interface (2) and *API caller* (2). The full list of papers and the annotations are available at https://bit.ly/2xNBZqF.

Since the stand-alone pattern represents instances of bots developed independently of other systems, and whose design is a reflection of the developer choices, not of the architectural layer or related artefacts, we focus the discussion below on the other seven patterns.

## In-app assistant

An *in-app assistant* is a bot that lives inside an existing application (e.g., a website or a desktop application) and extends the apps's features with conversational capabilities, e.g., using pop-ups, embedded components or conversational landing pages. An early example in this space is "Clippy", Microsoft's assistant for its desktop Office suite, although more recent work and frameworks focus on augmenting websites, e.g., to provide customer service.

The available user intents vary depending on the specific supporting platform, but they typically fall in one of the following categories: *question & answering* (conversational FAQ), *navigation support* (shortcuts and guidance in certain user workflows), *guided exploration* (assisting users in identifying a product or resource), *data collection* (a conversational form replacement), and *chit-chatting*. Chatbot actions can result in consulting the knowledge base to serve a request, triggering in-app navigation, or connecting to an integrated system to execute related tasks (e.g., post a new ticket). In terms of dialog control, one of the prominent aspects of this pattern is that the dialog context can be defined by the user actions in the app (e.g., current page or interaction history). Thus, the user can request or be prompted with assistance depending on the navigation context. As for the training, it can range from simply providing an URL in input (in the case of websites) and connecting to a knowledge base (e.g., a ticket system), to an explicit design of the conversation flow.

While we identified only few works in this pattern, the development of this type of bots is widely supported by commercial platforms. Among the research works, SuperAgent[6] is a prominent example of a chatbot extension for online shopping that can leverage both publicly available product information and user-generated data to support customer service efforts. Commercial solutions vary in the extent of support and automation in the chatbot generation. Acobot (https://acobot.ai) is an example of a platform where the chatbot is built from the content of the website and oriented toward assisting users during their browsing activity; while Instabot (https://www.instabot.io) facilitates domain experts in generating the chatbots but require the explicit definition of conversation flows.

## Conversational GUI agent

A *conversational GUI agent* is a bot that provides conversational access to the graphical user interface (GUI) of existing applications, much like traditional screen readers for blind or visually impaired users. For example, a bot in this category would allow a user to browse the contents of a news website using Amazon Alexa or interact with a mobile app using Google Assistant.

Intents in this pattern go from generic navigation (e.g., opening a website) and interacting with individual UI components (e.g., clicking a button), to high-level intents defined by the offerings of the specific application, be it transactional (e.g., booking a ticket) or informational (e.g., performing a query). Chatbot actions in this pattern mimic user interactions with UI components, and selectively fetch content so as to serve and respond to user requests. Dialog control is defined by the structure and state of the application. Training, from the perspective of the developer, consists in either augmenting the application definition with bot-specific annotations (and possibly utterances) so that the bot definition is part of the app, or defining conversation flows externally while referencing UI elements.

Approaches to Conversational UI agents can be found notably for Web browsing and mobile applications. Among the most recent proponents of this idea in web browsing, Baez, Daniel and Casati[7] propose full conversational access to websites where a chatbot mediates the interaction between the user and the website, allowing users to express their goals in natural language; while Ripa et al.[8] focus on making informational queries over content intensive websites accessible via voice-based interfaces (e.g., smart speakers). For mobile applications, Tarakji et al.[9] propose a framework that allows third-party developers to create voice user interfaces for existing Android apps, thus enabling users to interact with the apps in their mobile phones from smart speakers.

## Conversational API caller

A *conversational API caller* is a bot that is able to mediate between a user and a generic back-end service like a RESTful API or SOAP web service identified by the user. An example is a bot that is generated directly from an OpenAPI specification[10].

The intents in this case are a reflection of the functionality exposed by the API endpoint, which are derived from the API specification (e.g., OpenAPI, WSDL, WADL) or the service signature. Browsing and navigating a resource model (i.e., navigating through the relations between web artefacts) is conceptually another possibility in this pattern. Actions in this context refer to "external" calls to the associated service using HTTP / SOAP invocations. Dialog control is focused on *slot filling*, meaning collecting the necessary parameters for invoking the APIs. The input to the training and chatbot generation process consists in providing the API specification[10], or a service knowledge graph[11].

The generation of conversational interfaces from API descriptions is an approach that has been conceptualized and prototyped[10,11], but has not percolated in commercial products yet. Among the frameworks proposed in the literature, Varizi et al.[10] turns an OpenAPI specification directly into a chatbot implementation, although with limited NLU capabilities based on conventions. Zamanirad[12] instead proposes a system that allows bot developers to simply state the goal of the bot or an example utterance, identify a matching API from an evolving knowledge graph of services, and generate the code for a target chatbot platform.

## Conversational business process interface

A *conversational business process interface* is a chatbot that enables human process participants to interact with a business process in natural language, where the business process may orchestrate multiple human and software agents (APIs). Examples of business processes studied in literature are alert management[12] or IT change management[13].

Typical intents in interactions with a business process are obtaining information about the structure of the process (e.g., actors, tasks, inputs and outputs) or about the progress of a process in execution (e.g., which task is currently being processed, who is responsible for it), or performing activities (tasks) to advance the state of the process. Actions enact queries to obtain information about the process, and API calls to advance it. Dialog control is driven by the model and state of the process. The input to the training is the process model definition (e.g., BPMN).

One of the first evidence we found of conversational process interfaces is a patent by Google[12], which proposes the use of a chatbot component in so-called communication-enabled business process (CEBP) applications, i.e., applications able to orchestrate reactive and proactive communication events; no specific details about the internals of the chatbot infrastructure are however provided. Kalia et al.[13] propose a methodology for the extraction of a bot from a BPMN business process model with the goal of automating the process and providing process participants with a conversational UI; an IBM Watson model for the chatbot is constructed manually.

## Conversational IoT interface

A *conversational IoT interface* is a chatbot that provides conversational access to one or more physical devices to read device properties and/or enact actions through the devices. Examples are voice interfaces for smart homes or voice controls for vehicles or robots.

The intents and actions of the bots that implement this pattern are mostly limited to the specific devices' capabilities, such as reading a temperature measure or opening the window blinds; small talk intents without specific effect on the devices are of lower importance, if supported at all. Dialog control primarily focuses on command interpretation and slot filling and does not require sophisticated internal logics; most ad-hoc implementations not based on pre-canned, AI-based frameworks adopt simple rule-based input interpretation. The training prevalently follows an ad-hoc methodology in function of the available device functionalities.

In terms of development support, RedBot (http://red-bot.io) is a chatbot platform for the development of chatbots for Node-RED IoT applications. The accompanying visual tool extends Node-RED's modeling language with chatbot-specific elements compatible with, among others, Telegram, Facebook Messenger, Alexa, Viber. Hidalgo-Paniagua et al.[14] extended RedBot to support controlling more complex physical robots. Einarsson et al.[15] propose SmartHomeML, a domain-specific modelling language for smart home applications, that allows users to easily define new skills (functionalities) that can then automatically be integrated into Amazon Alexa or Google Home.

## Conversational query engine

A bot is a *conversational query engine* if it provides conversational access to a (semi)structured data/knowledge base, that is, if it allows users to interrogate the schema (metadata) of a database and to inspect specific instances (data) by translating natural language instructions into low-level query languages, such as SQL or SPARQL.

Intents in this pattern are defined by the capabilities of the underlying query engine and associated language, although recent work also proposes to augment data exploration with statistical analyzes over the data (e.g., navigating data clusters)[16]. Actions are translated from user requests into engine-specific queries (e.g., SQL or SPARQL statements) or higher-level data processing functions, but the dialog control is ultimately data driven. In terms of the input to training and generation, the salient approaches rely on the database structure and contents, although some approaches are augmented by domain-specific[17].

Recent approaches propose techniques for generating conversational interfaces for relational databases given annotated database schemas[18], translating natural language requests into SPARQL queries[17], and providing conversational data exploration augmented by statistical properties in the data, enabling users to navigate data clusters in guided dialogs[16].

## Conversational information retriever

A *conversational information retriever* bot is one that enables natural language queries over a typically unstructured set of documents or data. Bots in this space are emerging as efficient tools to answer even complex and open questions to large and disperse data.

Intents in this pattern are described by the questions that users can ask, which are defined by the content and structure of the documents. We identify three main types of intents in this space: *question and answering*, typically questions that can be answered by referring directly to the contents of a document (e.g., "What are the steps to preparing a sponge cake?"); *search & recommendation*, where the user engages in a conversation with a bot to search and discover relevant information or documents (e.g., "I'm looking for good sponge cake recipes"); and *document-centered queries*, aiming at inquiring about the metadata of documents ("When was this document updated?"). Actions in this pattern are essentially calls to an internal answer generation engine, which predicts the answer to the user's question based on the type of request. Regarding the dialog style, it is user initiated for when the system reacts to user queries but can also be proactive when providing recommendations. Queries can be answered either in a single turn or in multi-turn conversations. The training process starts with documents such as, FAQ pages, product manuals and spreadsheets.

Commercial platforms typically focus on supporting Q&A tasks, enabling customers to go from data to bot in minutes (e.g., https://www.qnamaker.ai, https://passage.ai) without requiring coding experience. These platforms often allow customers to import documents to create a knowledge base, modify the knowledge base and to customize the inferred conversation design. Recent works also propose frameworks to address specific domains or tasks, such as recommendation[19].

# DISCUSSION AND OUTLOOK

In this article we brought attention to the problem of chatbot integration as an emerging area of research. The eight patterns we describe in this article show that integrating conversational capabilities into existing software systems comes in very diverse flavors, depending on which type of service the target chatbot should deliver and on where in a system's architecture it wants to source conversational knowledge from. We highlighted how these patterns inform the design and capabilities of chatbots, while providing relevant pointers, as a first step towards further investigating associated integration challenges. We highlight the key differences between the identified patterns in Table 1.

Table 1 - Summary of chatbot integration patterns and main dimensions

| | Intents | Training / config. | Dialog control | Actions | Frameworks |
|---|---|---|---|---|---|
| **Stand-alone agent** | Generic and defined by developers | Ad hoc identified by developer | Custom or ready framework | Custom developed by developers | Rasa, DialogFlow, IBM Watson |
| **In-app assistant** | Contextual Q&A, direct navigation, data input, guided exploration, chit-chat | App content and structure, underlying KB, domain-specific models | Hybrid generic and GUI-driven, explicit conversation flows for guidance | App navigation, KB interrogation for Q&A, contextual help, transaction execution | Acobot, Instabot |
| **GUI agent** | Generic app navigation, data input, app-specific functions | App content and structure, external training data, pre-trained models for specific UI actions | GUI-driven (state depends on GUI), exploratory | Mimic user interactions, orient user inside app, read out content | -- |
| **API Caller** | API access or exploration, resource exploration | API specification, sample data, reuse training data from similar APIs | Driven by interaction protocol of API, exploratory | Issue HTTP / SOAP calls, visualize data | -- |
| **Business process interface** | Obtain process model information or process status updates, execute activities | Business process model (e.g., BPMN), pre-trained domain models | Process model driven, process state driven | Query model for activities, roles, actors, responsibilities, issue API calls | -- |
| **IoT interface** | Obtain device info, operate devices, automate operation | Device capabilities, device properties, pre-trained models | Device/environment status, command interpretation | Sense and actuate using HTTP / CoAP calls | RedBot |
| **Query engine** | Query metadata, traverse data schema, query data instances, obtain statistical analyses | Database schema, data instances, domain-specific ontologies | Data structure driven, iterative query construction, data analysis dependent | Issue queries (e.g., in SQL/SPARQL), apply data analysis functions, visualize results | -- |
| **Information retriever** | Generic Q&A, generic search, recommendation | Document content, data pool turned into a KB / KG | Mostly Q&A resolution, KB and topic driven; Explicit follow up | Guessing answers from KB / KG | QnA Maker, Passage.ai |

While the presence of these patterns shows efforts in enabling conversational access at all the three levels of the reference architecture, some patterns are still underdeveloped. Except for stand-alone agents, in-app assistants and conversational information retrievers, we however register a general lack of pattern-specific development aids. This comes somewhat as a surprise if we consider the wealth of use cases that ask for conversational capabilities that comply with the identified patterns. Just to mention few:

- *Fast prototyping*: all organizations today have their own websites for external or internal use. Conversational GUI agents or API callers could leverage on these resources for the fast implementation of conversational services.
- *Supporting domain-experts in analytical tasks*, by generating conversational interfaces to domain-specific databases or datasets, either with query engines or information retrievers.
- *Improving accessibility and ubiquitous access*, by providing conversational access to applications via Conversational GUIs.
- *Supporting automation and domain-specific workflows*, e.g., by replacing repetitive tasks with bots that can collaborate with humans, leveraging on Conversational GUI and BPs.

Raising the need for development assistance seems timely and wants to stimulate research. We identified patterns and solutions developers can already rely on but designing effective conversational interactions with existing systems is still in its infancy[20]. As the space of interconnected devices develop, we are likely to see new patterns emerge, as indicated by vision papers on conversational interfaces to drones and self-driving cars.

**Limitations:** The integration patterns described in this work are tied to the three-tier architecture of distributed systems, and the pattern definition adopted. They represent the most salient patterns from our analysis of 347 papers reporting on chatbot systems, although the list cannot be considered exhaustive due to the risk on missing out relevant work.

# REFERENCES

1. Baez, M., Casati, F., Gaedke, M., & Dustdar, S. (2020). Remembering Florian Daniel. IEEE Internet Computing, 24(03), 58-59.
2. Harms, J. G., Kucherbaev, P., Bozzon, A., & Houben, G. J. (2018). Approaches for Dialog Management in Conversational Agents. *IEEE Internet Computing*, *23*(2), 13-22.
3. Hussain, S., Sianaki, O. A., & Ababneh, N. (2019, March). A Survey on Conversational Agents/Chatbots Classification and Design Techniques. In *Workshops of AINA* (pp. 946-956). Springer, Cham.
4. F. Daniel, J. Yu, B. Benatallah, F. Casati, M. Matera and R. Saint-Paul. Understanding UI Integration: A survey of problems, technologies and opportunities. *IEEE Internet Computing 11(3)*, May/June 2007, Pages 59-66.
5. van der Aalst, W.M.P., Bichler, M. & Heinzl, A. Robotic Process Automation. Bus Inf Syst Eng 60, 269–272 (2018). https://doi.org/10.1007/s12599-018-0542-4
6. Cui, Lei, Shaohan Huang, Furu Wei, Chuanqi Tan, Chaoqun Duan, and Ming Zhou. Superagent: A customer service chatbot for e-commerce websites. *Proceedings of ACL 2017*, *System Demonstrations* (2017): 97-102.

7. Baez, M., Daniel, F., & Casati, F. (2019, November). Conversational Web Interaction: Proposal of a Dialog-Based Natural Language Interaction Paradigm for the Web. In *International Workshop on Chatbot Research and Design* (pp. 94-110). Springer, Cham.
8. Ripa, G., Torre, M., Firmenich, S., & Rossi, G. (2019, July). End-User Development of Voice User Interfaces Based on Web Content. In *International Symposium on End User Development* (pp. 34-50). Springer, Cham.
9. Tarakji, A. B., Xu, J., Colmenares, J. A., & Mohomed, I. (2018, June). Voice enabling mobile applications with UIVoice. In *Proceedings of the 1st International Workshop on Edge Systems, Analytics and Networking* (pp. 49-54).
10. Vaziri, M., Mandel, L., Shinnar, A., Siméon, J., & Hirzel, M. (2017, October). Generating chat bots from web API specifications. In *Proceedings of the 2017 ACM SIGPLAN International Symposium on New Ideas, New Paradigms, and Reflections on Programming and Software* (pp. 44-57).
11. Zamanirad, Shayan. (2019). Superimposition of Natural Language Conversations over Software Enabled Services.
12. Gaulke, D. and Kornbluh, D., Avaya Inc, 2015. Interactive user interface to communication-enabled business process platforms method and apparatus. U.S. Patent 9,043,407.
13. Rizk, Y., Bhandwalder, A., Boag, S., Chakraborti, T., Isahagian, V., Khazaeni, Y., Pollock, F. and Unuvar, M., 2020. A Unified Conversational Assistant Framework for Business Process Automation. arXiv preprint arXiv:2001.03543.
14. Hidalgo-Paniagua, Alejandro, Andrés Millan-Alcaide, Juan P. Bandera, and Antonio Bandera. "Integration of the Alexa assistant as a Voice Interface for Robotics Platforms." In *Iberian Robotics conference*, pp. 575-586. Springer, Cham, 2019.
15. Einarsson, Atli F., Patrekur Patreksson, Mohammad Hamdaqa, and Abdelwahab Hamou-Lhadj. "SmarthomeML: Towards a domain-specific modeling language for creating smart home applications." In *2017 IEEE International Congress on Internet of Things (ICIOT)*, pp. 82-88. IEEE, 2017.
16. Sellam, T., & Kersten, M. (2016, June). Have a chat with clustine, conversational engine to query large tables. In *Proceedings of the Workshop on Human-In-the-Loop Data Analytics* (pp. 1-6).
17. Rajosoa, M., Hantach, R., Abbes, S. B., & Calvez, P. (2019). Hybrid Question Answering System based on Natural Language Processing and SPARQL Query.
18. Castaldo, Nicola, Florian Daniel, Maristella Matera, and Vittorio Zaccaria. Conversational Data Exploration. In *International Conference on Web Engineering*, pp. 490-497. Springer, Cham, 2019.
19. Iovine, A., Narducci, F., & Semeraro, G. (2020). Conversational Recommender Systems and natural language: A study through the ConveRSE framework. *Decision Support Systems*, 113250.
20. Mavridis, P., Huang, O., Qiu, S., Gadiraju, U., & Bozzon, A. (2019, June). Chatterbox: Conversational interfaces for microtask crowdsourcing. In Proceedings of the 27th ACM Conference on User Modeling, Adaptation and Personalization (pp. 243-251).